\documentclass[12pt,a4paper]{article}

\usepackage{amssymb}
\usepackage{amsmath}
\usepackage{graphicx}
 
\newcommand{\be}{\begin{equation}}
\newcommand{\ee}{\end{equation}}
\newcommand{\bea}{\begin{eqnarray}}
\newcommand{\eea}{\end{eqnarray}}
\newcommand{\lb}{\label}
\newcommand{\bdm}{\begin{displaymath}}
\newcommand{\edm}{\end{displaymath}}

\begin{document}

\begin{titlepage}

\noindent
\begin{center}
\vspace*{1cm}

{\large\bf CAN EFFECTS OF QUANTUM GRAVITY BE OBSERVED IN THE COSMIC
  MICROWAVE BACKGROUND?} 

\vskip 1cm

{\bf Claus Kiefer}\footnote{\tt kiefer@thp.uni-koeln.de} {\bf and
  Manuel Kr\"amer}\footnote{\tt mk@thp.uni-koeln.de}   
\vskip 0.4cm
Institute for Theoretical Physics,\\ University of Cologne, \\
Z\"ulpicher Stra\ss e~77,
50937 K\"oln, Germany.\\ {\tt http://www.thp.uni-koeln.de/gravitation/}
\vspace{1cm}

\begin{abstract}
We investigate the question whether small quantum-gravitational
effects can be observed in the anisotropy spectrum of the cosmic
microwave background radiation. An observation of such an effect is
needed in order to discriminate between different approaches to quantum
gravity. Using canonical quantum gravity with the Wheeler--DeWitt
equation, we find a suppression of
power at large scales. Current observations only lead to an upper
bound on the energy scale of inflation, but the framework is general
enough to study other situations in which such effects might indeed be seen.
\end{abstract}

\vskip 2cm

{\bf Essay awarded first prize in the }\\ {\bf Gravity Research
  Foundation essay competition 2012}

\end{center}

\end{titlepage}


Advances in physics often come with small effects. One famous example
is the perihelion precession of Mercury. After all the known
perturbations from other planets were taken into account, there
remained for the precession an amount of about $43''$ per century,
which could not be 
understood by the knowledge of the 19th century. It was one of
Einstein's great triumphs to find a natural explanation of this effect
from his theory of general relativity. Another important example is
the Lamb shift, which was discovered in 1947 by Lamb and Retherford. This shift
removes the degeneracy of the spectral lines $2s_{1/2}$ and $2p_{1/2}$
in the hydrogen atom and is a consequence of quantum electrodynamics (QED). 
The significance of this effect lies both in its accurate measurement and in
its theoretical derivation. For the latter, one needs the
sophisticated methods of renormalization, which at that time had only been newly
developed. 
One thus cannot overemphasize the crucial role that the
Lamb shift plays for quantum field theory in general and QED in
particular. This is reflected in the following words 
that Dyson wrote to Lamb on the occasion of his 65th birthday \cite{Lamb}:

\begin{quote}
Those years, when the Lamb shift was the central theme of physics,
were golden years for all the physicists of my generation. You were
the first to see that that tiny shift, so elusive and hard to measure,
would clarify in a fundamental way our thinking 
about particles and fields.
\end{quote}

In our essay, we shall address the question whether small effects can
also guide us in the search for a quantum theory of gravity. 
It is notoriously difficult to construct a consistent quantum theory
of gravity \cite{oup}. The direct
quantization of general relativity leads to a non-renormalizable
theory at the perturbative level. Non-perturbative approaches are available,
but they do not yet exist in a complete form. It is thus of the utmost
importance to look for observational hints that could play the role
the Lamb shift played for the development of QED. 

The reason why effects of quantum gravity have not yet been seen lies
in their extreme smallness. One expects that these effects are suppressed by
a factor proportional to $(E/m_{\rm P})^2$, where $E$ is the respective
energy scale and $m_{\rm P}$ is the Planck mass, which we define here as
$m_{\rm P}=\sqrt{3\pi\hbar c/2G}\approx 2.65 \times 10^{19}$ GeV
for later convenience. (In the following, we set $\hbar=c=1$.)
Even for energies as high as the ones available
at the Large Hadron Collider (LHC), $E\approx 1.4\times 10^4\, {\rm
  GeV} $, possible effects of quantum 
gravity are far too small to be seen.

In the following, we address the question whether effects
of quantum gravity can be seen in cosmological observations.
We use canonical quantum gravity in its geometrodynamical form. 
 This framework can be motivated by
following the heuristic route that led Erwin Schr\"odinger 
to his famous wave equation in 1926. What he did was to formulate
classical mechanics in Hamilton--Jacobi form and to `guess' a wave
equation that leads to the Hamilton--Jacobi equation in what we
now call the semiclassical or WKB limit. The same can be done for general
relativity \cite{oup,essay}. One can transform Einstein's equations
into Hamilton--Jacobi form and `guess' the wave equation that
leads back to it in the semiclassical limit. The result is the
Wheeler--DeWitt equation, which has the form
\be
\lb{wdw}
{\mathcal H}\Psi=0,
\ee
where ${\mathcal H}$ denotes the total Hamiltonian of gravity and of all
non-gravitational degrees of freedom. It is important to note that
this equation is timeless, that is, it does not contain any external
time parameter \cite{essay}. Even though the Wheeler--DeWitt equation
might not hold at the most fundamental level, the point here is that it
should be valid at least as an effective equation, because it gives
the correct semiclassical limit.

Here, we apply this equation to an inflationary model of the
early universe and investigate whether one can derive effects from it that
are potentially observable in the anisotropy spectrum of the cosmic
microwave background (CMB) radiation \cite{KK12}. 

We consider a flat Friedmann--Lema\^itre universe 
with scale factor $a\equiv\exp(\alpha)$ and a scalar field $\phi$ 
with potential $\mathcal{V}(\phi)$, which
plays the role of the inflaton. 
We assume that classically the standard slow-roll condition of the
form $\dot{\phi}^2 \ll \vert\mathcal{V}(\phi)\vert$ holds.
For definiteness, we choose the simple potential 
$
\mathcal{V}(\phi) = \frac{1}{2}\,m^2\phi^2,
$
but any other potential will be fine as long as the slow-roll condition is
satisfied. 

In order to describe the anisotropies of the CMB, additional degrees
of freedom describing small fluctuations must be introduced. These
correspond to density fluctuations and small gravitational waves.  
Here, we restrict ourselves to the fluctuations of the scalar field,
but the extension to the general case is straightforward. Since the
fluctuations are small, one can neglect their coupling and treat the
various wavenumbers $k$ as independent. One can then formulate the
Wheeler--DeWitt equation \eqref{wdw} for the wave function
$\Psi(\alpha,\phi,\{f_k\})$, where the $f_k$ denote the Fourier
components of the $\phi$-fluctuations (also called
modes) \cite{Hall85}. The slow-roll 
approximation is implemented in the quantum
theory by demanding that the $\phi$-derivatives of the wave function
be much smaller than the $\alpha$-derivatives. 
The smallness of
the fluctuations allows us to write $\Psi$ as a product of functions
$\Psi_k(\alpha,f_k)$, which can be studied independently.

Since one is interested in quantum-gravitational corrections to
standard formulae, it is sufficient to solve the Wheeler--DeWitt
equation in a Born--Oppenheimer type of approximation
\cite{oup,essay,Hall85}. This scheme is well known from molecular
physics and is based on the idea that the degrees of freedom can be
separated into `fast' and `slow' ones. In the case of molecules, the
slow ones are the nuclei, and the fast ones are the electrons. In
the present case, the slow variable is $\alpha$, and
the fast variables are the fluctuations $f_k$.

The Born--Oppenheimer expansion is implemented by an expansion with
respect to the inverse Planck mass squared. 
The order $\mathcal{O}(m_{\rm P}^0)$
corresponds to the limit of quantum theory in an external
background. An approximate time parameter $t$ is then at our disposal,
and the wave functions for the fluctuations obey an approximate
Schr\"odinger equation with respect to $t$.
At this order, one obtains the standard results for 
quantum fluctuations in an inflationary universe. 
If one assumes that the fluctuations start in their ground states,
they evolve into squeezed states when
during inflation their wavelengths become larger than the 
Hubble radius $H^{-1}$. The power spectrum of these fluctuations is
then calculated when they re-enter the Hubble radius during the
radiation-dominated phase. It turns out that this spectrum is
approximately scale-invariant. The fluctuations then lead to the 
anisotropy spectrum of the CMB, whose approximate scale invariance
has been confirmed by observation \cite{Koma11}. 
 
In order to calculate quantum-gravitational modifications,
one must go beyond the order $\mathcal{O}(m_{\rm P}^0)$ in
\eqref{wdw}. It has been shown for the full Wheeler--DeWitt equation
that the next order,  
$\mathcal{O}(m_{\rm P}^{-2})$, leads to quantum-gravitational
corrections to the Schr\"odinger equation that are proportional to
$m_{\rm P}^{-2}$ \cite{KS91}. 
In the present case, this leads to a corrected Schr\"odinger equation
of the form \cite{KK12}
\be \label{corr_Schr_eq}
\text{i}\,\frac{\partial}{\partial t}\,\psi^{(1)}_{k} \approx
\mathcal{H}_{k}\psi^{(1)}_{k}- 
\frac{1}{2{\rm e}^{3\alpha}m_{\rm P}^2H^2}\frac{\bigl(
\mathcal{H}_{k}\bigr)^2\psi^{(0)}_{k}}{\psi^{(0)}_k}  \psi^{(1)}_k , 
\ee
where the $\psi^{(0)}_{k}$ and the $\psi^{(1)}_{k}$ denote the wave
functions of the 
fluctuations at $\mathcal{O}(m_{\rm P}^0)$ and $\mathcal{O}(m_{\rm
  P}^{-2})$, respectively, and $\mathcal{H}_{k}$ is the Hamiltonian for
the modes.  One can now calculate the modification of the power
spectrum caused by the correction term in \eqref{corr_Schr_eq}
\cite{KK12}. 

Denoting the original power spectrum and the corrected spectrum by 
$\Delta_{(0)}^2(k)$ and $\Delta_{(1)}^2(k)$, respectively, one finds 
$
\Delta_{(1)}^2(k) = \Delta_{(0)}^2(k)\,C_k^2
$
with 
\begin{equation}
\label{Ck}
C_k\approx\left(1 -
    \frac{43.56}{k^3}\,\frac{H^2}{m_{\text{P}}^2}\right)^{\!\!-\frac{3}{2}}
  \!\!\left(1 
  - \frac{189.18}{k^3}\,\frac{H^2}{m_{\text{P}}^2}\right).
\end{equation}
Scale invariance is thus broken at this level.
The size of the corrections is given by $(H/m_{\rm P})^2$.
 This ratio is expected to be much higher than the size of
quantum-gravitational corrections in the laboratory. 

The correction function $C_k$ is displayed in Figure~1 for the special
value $H=10^{14} \ {\rm GeV}$.
At large $k$ (small scales), it approaches one, but it decreases
monotonically to zero for small $k$ (large
scales). Quantum-gravitational effects are thus most prominent at
large scales. This is not surprising, since these scales are the
earliest to leave the Hubble radius during inflation. 
However, the whole approximation scheme breaks down when $C_k$
approaches zero.

\begin{figure}[h]
\begin{center}
  \includegraphics[width=0.75\textwidth]{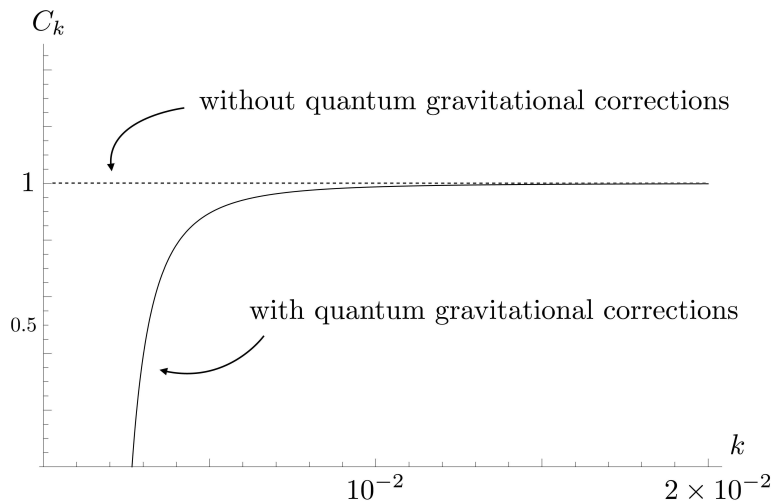} 
\end{center}
 \caption[]{The function $C_k$ for $H=10^{14}\ {\rm GeV}$.}
\end{figure}

One therefore gets a concrete prediction from a conservative approach to
quantum gravity: there must be a {\em suppression of power at large
scales}. At present, there is no unambiguous observation of such an
effect \cite{Koma11}. Therefore, one only gets an upper bound on
the Hubble parameter $H$ \cite{KK12}, 
 \begin{equation}
\label{Hbound}
H \lesssim 1.4\times 10^{-2}\,m_{\text{P}} \approx
4\times10^{17}\,\text{GeV}.
\end{equation}
This is, in fact, a weaker bound than the 
observational bound from the tensor-to-scalar ratio of the CMB fluctuations,
which gives $H\lesssim 10^{-5}m_{\text{P}}\sim
10^{14}\,\text{GeV}$ (see, e.g., \cite{Baumann}); 
therefore, we have chosen the value $10^{14}\,\text{GeV}$ 
in Figure~1.
 Nevertheless, this investigation opens
up a window to an energy range where quantum-gravitational effects can
become large enough to be observable, in contrast to the situations
hitherto studied. 

One can also compare the above results with the predictions from other
approaches to quantum gravity. Non-commutative geometry and effects
from string theory also lead to a suppression of power at large
scales, although of a different type \cite{suppression}.
This is not the case for loop quantum cosmology -- there 
one finds an enhancement of power at large scales \cite{BCT}. 

A comparison of these results
could give us the unique opportunity to discriminate
observationally between different approaches. To paraphrase Dyson's
words from above, the discovery of
a small quantum-gravitational effect would clarify in a
fundamental way our thinking 
about gravity, particles, and fields.



\end{document}